\documentclass[a4paper,12pt]{article}
\usepackage{tikz}
\usepackage[english]{babel}
\usepackage[utf8x]{inputenc}
\usepackage{amssymb,amsmath,graphicx,ucs,ifthen}

\def\Xint#1{\mathchoice
   {\XXint\displaystyle\textstyle{#1}}%
   {\XXint\textstyle\scriptstyle{#1}}%
   {\XXint\scriptstyle\scriptscriptstyle{#1}}%
   {\XXint\scriptscriptstyle\scriptscriptstyle{#1}}%
   \!\int}
\def\XXint#1#2#3{{\setbox0=\hbox{$#1{#2#3}{\int}$}
     \vcenter{\hbox{$#2#3$}}\kern-.5\wd0}}

\def\dashint{\Xint-}

\newcommand{\eq}{\begin{equation}}
\newcommand{\eqx}{\end{equation}}
\newcommand{\eqs}{\begin{equation*}}
\newcommand{\eqsx}{\end{equation*}}
\newcommand{\eqn}{\begin{eqnarray}}
\newcommand{\eqnx}{\end{eqnarray}}
\newcommand{\eqns}{\begin{eqnarray*}}
\newcommand{\eqnsx}{\end{eqnarray*}}
\newcommand{\alg}{\begin{align}}
\newcommand{\algx}{\end{align}}

\newcommand{\f}[2]{\frac{#1}{#2}}

\newcommand{\lm}{\lambda}
\newcommand{\Lm}{\Lambda}

\newcommand{\sg}{\sigma}

\newcommand{\bt}{\beta}

\newcommand{\gm}{\gamma}

\newcommand{\qqqq}{\quad\quad\quad\quad}
\newcommand{\qq}{\quad\quad}

\newcommand{\nn}{{\cal N}}

\newcommand{\slii}{\ensuremath{\mathfrak{sl}(2)}\ }
\newcommand{\slc}{\ensuremath{\mathfrak{sl}(2,\mathbb{C})}}

\newcommand{\jc}{j_{cl}}
\newcommand{\nuc}{\nu_{cl}}

\newcommand \thetitle {Approaching the BFKL pomeron via integrable classical solutions} %%%%%%%%%%%%%% ?

\title{\thetitle}

\author{Romuald A.\ Janik\thanks{e-mail: \href{mailto:romuald@th.if.uj.edu.pl}{\tt romuald@th.if.uj.edu.pl}}
   \and Paweł Laskoś-Grabowski\thanks{e-mail: \href{mailto:plg@th.if.uj.edu.pl}{\tt plg@th.if.uj.edu.pl}}\\
   }

\date{Institute of Physics, Jagiellonian University\\ul. Reymonta 4, 30-059 Kraków, Poland}

\DeclareMathOperator \cn {cn}

\DeclareMathOperator \IM {Im}
\DeclareMathOperator \disc {disc}

\newcommand \figref [1] {Fig.\ \ref{#1}}

\PrerenderUnicode{łś}
\usepackage[unicode=true,pdftitle={\thetitle},pdfauthor={Romuald A. Janik, Paweł Laskoś-Grabowski},pdftex]{hyperref}
\begin{document}

\maketitle

\begin{abstract}
We identify classical string solutions which directly give the classical part of the
strong coupling pomeron intercept. The relevant solution is a close cousin
of the GKP folded string, which is not surprising given the known relation with
twist-2 operators. Our methods are applicable, however, also for nonzero 
conformal spin where we do not have a clear link with anomalous dimensions
of a concrete class of operators.
We analyze the BFKL folded string from the algebraic curve perspective and
investigate its possible particle content.
\end{abstract}

\thispagestyle{empty}\newpage

\section{Introduction}

The understanding of high energy scattering in gauge theories is both very important
and interesting. Of particular interest is the specific case of the Regge limit, namely
the limit of high energy and fixed momentum transfer. The relevant dynamics in QCD
depends on the magnitude of the momentum transfer in comparision with $\Lm_{QCD}$.
When it is small, we are dealing with the physics of the `soft pomeron' which
is nonperturbative and still remains mysterious from the theoretical point of view.
When the momentum transfer is large enough (but fixed), perturbative computations
are applicable and the description is in terms of the `hard pomeron:'
the BFKL pomeron~\cite{BFKL}.
This regime is relevant for applications to Deep Inelastic Scattering processes.

The BFKL pomeron leads to power law behaviour of scattering amplitudes with energy $s^j$,
where $j$ is the pomeron intercept.
Its value depends on quantum numbers of the particular state of the pomeron being exchanged. 
In the Regge limit, there is
a clear separation between the longitudinal plane defined by the collision and
the transverse plane. The chief characteristic of the BFKL equation is its invariance
under the \slc{} symmetry group of the transverse plane. Therefore the intercept
is always a function of the relevant principal continuous series representation
of \slc{} parametrized by\footnote{In case of compound states of more than two reggeized
gluons, additional discrete quantum numbers will appear.}
\eq
h=\f{1+n}{2}+i \nu \qq \tilde{h}=\f{1-n}{2}+i\nu
\eqx

Currently, the BFKL pomeron intercept is known at leading and next-to-leading order both in QCD
and in $\nn=4$ SYM \cite{NLO}. In fact at leading order the results exactly coincide in the two theories.
At NLO level, differences appear but still there are many similarities in the structure
of the expressions,
hence insight into the properties of the BFKL pomeron in $\nn=4$ SYM at any coupling
would be also significant for QCD physics.

The reason why it is interesting to study the BFKL pomeron in $\nn=4$ SYM is that due to
the AdS/CFT correspondence we can access nonperturbative strong coupling physics.
In particular, now we know the BFKL pomeron properties at strong coupling, 
where the intercept starts off at 2 \cite{JP} (corresponding to graviton exchange in the dual
picture) and decreases as we lower the coupling \cite{BPST,Costa,KL}.    %%%%%% = BPS ?
Even more so, using the methods of integrability in the AdS/CFT correspondence,
we may hope to obtain an exact expression for the intercept valid at any coupling which
interpolates between the known LO and NLO BFKL and the strong coupling
behaviour. However, we are still away from realizing this goal.

Currently the dominant approach in tackling the problem of the BFKL pomeron using
integrability is
to employ a relation between the analytical continuation of the anomalous dimensions
of twist-2 operators and the BFKL intercept. This relation has been initially used
to derive constraints on the higher loop anomalous dimensions \cite{KLRSV,TWIST2L4,TWIST2L5}
and to show the neccessity of wrapping corrections. However now, as our knowledge
of the anomalous dimensions is much more complete, it has been exploited 
in the opposite direction to
gain knowledge about the strong coupling BFKL intercept \cite{Costa,KL}.
The drawback of this way of approach is that we are led to study 
\emph{the analytical continuation} of the anomalous dimensions 
at nonphysical values of the spin. This poses significant challenges for
the current exact formulations of the $\nn=4$ SYM spectral problem which typically
provide just numerical values (in case of generic coupling) of the anomalous
dimensions for \emph{integer} spin.
However, some work in this direction has already been carried out \cite{RJBAXTER}.

In this paper we would like to adopt a different, more direct approach,
where we would like to identify string configurations relevant for BFKL
without assuming \emph{a~priori} any analytical continuation properties.
This is especially interesting in view of a more general situation,
e.g.\ with nonzero values of the conformal spin, where we lack a clear picture
of what would be the relevant class of analytically continued operators.
Similarly, apart from the BFKL pomeron, in perturbative gauge theory there is
a whole range of compound states of more than two reggeized gluons which
also give contributions to scattering amplitudes. It is not known what would
be their behaviour/counterparts at strong coupling. We hope that the direct
approach advocated in this paper might shed some light on these issues.

In this paper we will restrict ourselves to the strong coupling regime.
Let us first review the setting of the spectral problem in this context.

The spectral problem amounts to expressing the energy (equivalently, the
dimension) as a function of all other conserved charges of the theory, i.e.\ 
two spins and three R-charges, and other quantum numbers labelling different
states with the same conserved charges. At strong coupling, an important
class of states corresponds to classical spinning string solutions
in $AdS_5 \times S^5$. These solutions have all conserved charges proportional
to $\sqrt{\lm}$ and a specific solution to the equations of motion
of the $AdS_5 \times S^5$ $\sg$-model leads to an explicit
relation
\eq
\f{E}{\sqrt{\lm}} = F\left( \f{S_1}{\sqrt{\lm}}, \f{S_2}{\sqrt{\lm}};
\f{J_1}{\sqrt{\lm}},\f{J_2}{\sqrt{\lm}},\f{J_3}{\sqrt{\lm}} \right)
\eqx

We would like to adopt the same philosophy and find string configurations
which would similarly provide a relation between the strong coupling intercept
and the relevant quantum numbers. Again, in the classical limit we will
consider the case that all the quantum numbers scale like $\sqrt{\lm}$,
hence we are looking for a relation of the form
\eq
\f{j}{\sqrt{\lm}} =\tilde{F} \left( \f{\nu}{\sqrt{\lm}}, \f{n}{\sqrt{\lm}}, \ldots \right)
\eqx
This relation should follow from exhibiting a specific solution to the equations of motion
of the $AdS_5 \times S^5$ $\sg$-model.

One of the goals of the present paper is to identify these solutions and to analyze
them from the point of view of integrability of strings in $AdS_5 \times S^5$. Hence we
will identify the properties of the corresponding algebraic curve.
In particular we would like to analyze the particle interpretation of these solutions.
In this paper we will restrict ourselves to the case of zero conformal spin, in order
to have a direct counterpart of the more standard link with twist-2 operators.
However our methods are straightforward to apply for the general case of nonzero conformal
spin going beyond the twist-2 sector (using e.g.\ analogs of classical solutions 
in \cite{FT}).

The plan of this paper is as follows. In the next two sections we will review some basic facts
about weak and strong coupling BFKL and the identification of the pomeron quantum
numbers within the complexified conformal group.
Then we will move on to describe the relevant classical string solution and, after
reviewing for completeness the particle interpretation of algebraic curve constructions,
we will analyze this solution within the algebraic curve framework. We close the paper with
a summary and outlook.

\section{Scattering in the Regge limit and BFKL solution}

As stated in the introduction, the BFKL intercept is expressed as a function
of the \slc{} representations

\begin{align} h=\frac{1+n}2+i\nu \qquad \tilde{h}=\frac{1-n}2+i\nu \label{eq:sl2labels}\end{align}
where $n$ is an integer and $\nu$ is a continuous real parameter.

Often it is convenient to consider the case of vanishing conformal spin $n=0$ and
then use the notation $\gm=h=\tilde{h}=\f{1}{2}+i\nu$.
At weak coupling, we know the first two terms in the expression for the intercept:
\eq
j=1+\f{\lm}{4\pi^2} \left[ \chi_{LO}(\gm)+\f{\lm}{16\pi^2} \chi_\textit{NLO}(\gm)
+\ldots \right]
\eqx
Namely
\begin{align}
\chi_{LO}(\gm) &= 2\Psi(1)-\Psi(\gm)-\Psi(1-\gm) \\
\chi_\textit{NLO}(\gm) &= \Psi''(\gm)+\Psi''(1-\gm)+6\zeta_3 -2\zeta_2 \chi_{LO}(\gm)-2\Phi(\gm)-2\Psi(1-\gm) 
\end{align}
with
\eq
\Phi(\gm)=2\sum_{k=0}^\infty \f{(-1)^{k+1}}{k+\gm} \bt'(k+1)
\qq
\bt'(z) =\f{1}{4} \left[ \Psi'\left(\f{1+z}{2}\right) -\Psi'\left(\f{z}{2}\right)
\right]
\eqx

A general form of the strong coupling intercept was postulated to be
\eq
j(\nu)=2-\f{2+2\nu^2}{\sqrt{\lm}} \left(1+\sum_{n=2}^\infty 
\f{\tilde{j}_n(\nu^2)}{\lm^{(n-1)/2}}\right)
\eqx
with $\tilde{j}_n(\nu^2)$ being a polynomial of order $n-2$ \cite{Cornalba}.
Subsequently, it was argued in \cite{Costa}, that $\tilde{j}_n(\nu^2)$ should be 
a polynomial of order at most $n/2$.
This observation is in fact a direct consequence of the existence of an underlying
classical string solution. Namely, for a classical string solution both $j$ and $\nu$
should scale at most like $\sqrt{\lm}$. If there would be terms of higher order than
$n/2$ this would clearly be impossible.

At strong coupling we know the expression up to $\lm^{-2}$ terms, as well as the leading-$\nu$ term at $\lm^{-5/2}$~\cite{Costa,KL}
\begin{align}\label{e.strong}
j(\nu)&=2-\f{2+2\nu^2}{\sqrt{\lm}}\times\\\notag&\times\left(1+\frac{\tfrac12}{\sqrt\lm}+\frac{-\tfrac18+\tfrac32\nu^2}{\lm}+\frac{-1-3\zeta_3+(\tfrac{21}8-3\zeta_3)\nu^2}{\lm^{3/2}}+\frac{\frac{21}4\nu^4+O(\nu^2)}{\lambda^2}\right)
\end{align}

Let us now obtain the classical part of the strong coupling pomeron intercept 
result (\ref{e.strong}). We will introduce
\eq
\jc \equiv \f{j}{\sqrt{\lm}} \qq \nuc \equiv \f{\nu}{\sqrt{\lm}}
\eqx
Then the formula (\ref{e.strong}) gives
\eq
\jc=-2\nuc^2 -3\nuc^4- \f{21}{2} \nuc^6 + \ldots \label{eq:jcl}
\eqx
In the next two sections we will show how to reproduce this formula directly from classical string solutions and how to generate many more terms beyond those following from (\ref{e.strong}).

\section{The \slc{} subalgebra}

A crucial step in identifying solutions relevant to the BFKL Hamiltonian is the identification of an \slc{} subalgebra of the (complexified) conformal group~\cite{BST07,BC}:
\begin{align} J_0&=\tfrac12(-iD+M_{12}) &      J_+&=\tfrac12(P_1-iP_2) &      J_-&=\tfrac12(K_1+iK_2) \\
         \bar J_0&=\tfrac12(-iD-M_{12}) & \bar J_+&=\tfrac12(P_1+iP_2) & \bar J_-&=\tfrac12(K_1-iK_2)\end{align} 
where the directions 12 indicate the transverse plane.

One should now look for eigenstates of $J_0,\bar J_0$, which should correspond to the eigenvalues $h+m,\bar h+m'$, respectively, for some integer $m,m'$. By \eqref{eq:sl2labels}, this gives the following values for the charges:
\begin{align} -iD=1+2i\nu+m+m' \qquad M_{12}=n \end{align}

Another crucial identification, following from the relation between the BFKL Hamiltonian and the boost operator in the longitudinal plane, reads~\cite{BST07}
\begin{align} j=-iM_{+-}=-iM_{03}\end{align}
Therefore, one should look for solutions with nonzero $D,M_{03},M_{12}$. However, for simplicity we will restrict ourselves to the case of vanishing conformal spin $n=M_{12}=0$.

\section{Folded string solutions in  $AdS_3$}

\label{s.class}

The simplest example of a classical string solution with only two nonzero conserved charges
is the GKP folded string~\cite{GKP} which we briefly review below.
 
\subsubsection*{The GKP folded string}

We start from the solution for the GKP folded string  in $AdS_3$ subspace as parameterized in \cite{Ts10}, which in embedding coordinates reads
\begin{align} Y_0 &= \cosh\rho\sin\kappa t & Y_1 &= \sinh\rho\cos\omega t & Y_3&=0 \nonumber\\
              Y_5 &= \cosh\rho\cos\kappa t & Y_2 &= \sinh\rho\sin\omega t & Y_4&=0
\label{e.gkp}
\end{align}
and the Virasoro constraints relate the parameters $\kappa,\omega$ and function $\rho$ as
\begin{align} \omega = \frac2\pi K(k^2) \qquad \kappa = k\omega \qquad \sinh\rho = \frac k {\sqrt{1-k^2}} \cn \left(\omega\sigma+K(k^2) \middle\vert k^2 \right)
\end{align}

The only nonzero charges are equal to
\begin{align} E \equiv S_{50} &= \sqrt\lambda\frac2\pi \frac k {1-k^2} E(k^2) \label{eq:gkpE}\\
              S \equiv S_{12} &= \sqrt\lambda\frac2\pi \left( \frac1{1-k^2}E(k^2) - K(k^2) \right) \label{eq:gkpS}\end{align}
The physical range of $k$ is the interval $k \in(0,1)$ which interpolates between the limits of very short ($k\to0$) (small spin) and infintely long ($k\to1$) string (large spin).

\subsubsection*{The BFKL folded string}

As discussed above, we are interested in working with a configuration that has nonzero $S_{03},S_{54}$, and we can obtain one by reshuffling and complexifing the embedding 
coordinates with respect to (\ref{e.gkp})
\begin{align} Y_0 &=          -i \sinh\rho\sin\omega t & Y_3 &= \phantom{-i}\sinh\rho\cos\omega t & Y_1&=0 \nonumber\\
              Y_5 &= \phantom{-i}\cosh\rho\cos\kappa t & Y_4 &=          -i \cosh\rho\sin\kappa t & Y_2&=0\end{align}
In the above, the embedding condition ($\eta_{AB} Y^A Y^B=-1$) is preserved, and the Virasoro constraints take an identical form to the GKP case. We dub this solution the \emph{BFKL folded string}.

Now, the nonzero charges are
\begin{align} 2\nu = S_{54} &=          - i \sqrt\lambda\frac2\pi \frac k {1-k^2} E(k^2) \label{eq:S54}\\
                ij = S_{03} &= \phantom{-}i \sqrt\lambda\frac2\pi \left( \frac1{1-k^2}E(k^2) - K(k^2) \right) \label{eq:S03}\end{align}
where $k=\frac\kappa\omega$ just as before, but we no longer have the physical motivation to restrict this parameter to the real axis. Note that the form of the $S_{54}$ charge is consistent with the relation
\begin{align} L\cdot\tfrac12(P_0+K_0)\cdot L^{-1} = -iD \end{align}
in this case realised by $SO(4,2)$ rotations by $\frac\pi2$ in the embedding space in the planes 13, 02, and 04, or, in an unambiguous way, by
\begin{align} L = \exp \frac\pi2 \left(\frac4{3\sqrt3}(M_{42}+M_{20}+M_{40})+M_{31}\right)\end{align}
with $M_{ab}$ generators of the respective rotations.

The formulas \eqref{eq:S54}-\eqref{eq:S03} allow us to write out the classical intercept to a~priori arbitrary order, by expanding $k$ as a series in $\nuc$ and substituting it into the formula for $\jc$. The result
\begin{align}\jc=-2\nuc^2-3\nuc^4-\frac{21}2\nuc^6-\frac{391}8\nuc^8-\frac{8439}{32}\nuc^{10}+\cdots\end{align}
notably agrees with the expansion \eqref{eq:jcl}.

\begin{figure}[t]\includegraphics[width=\textwidth]{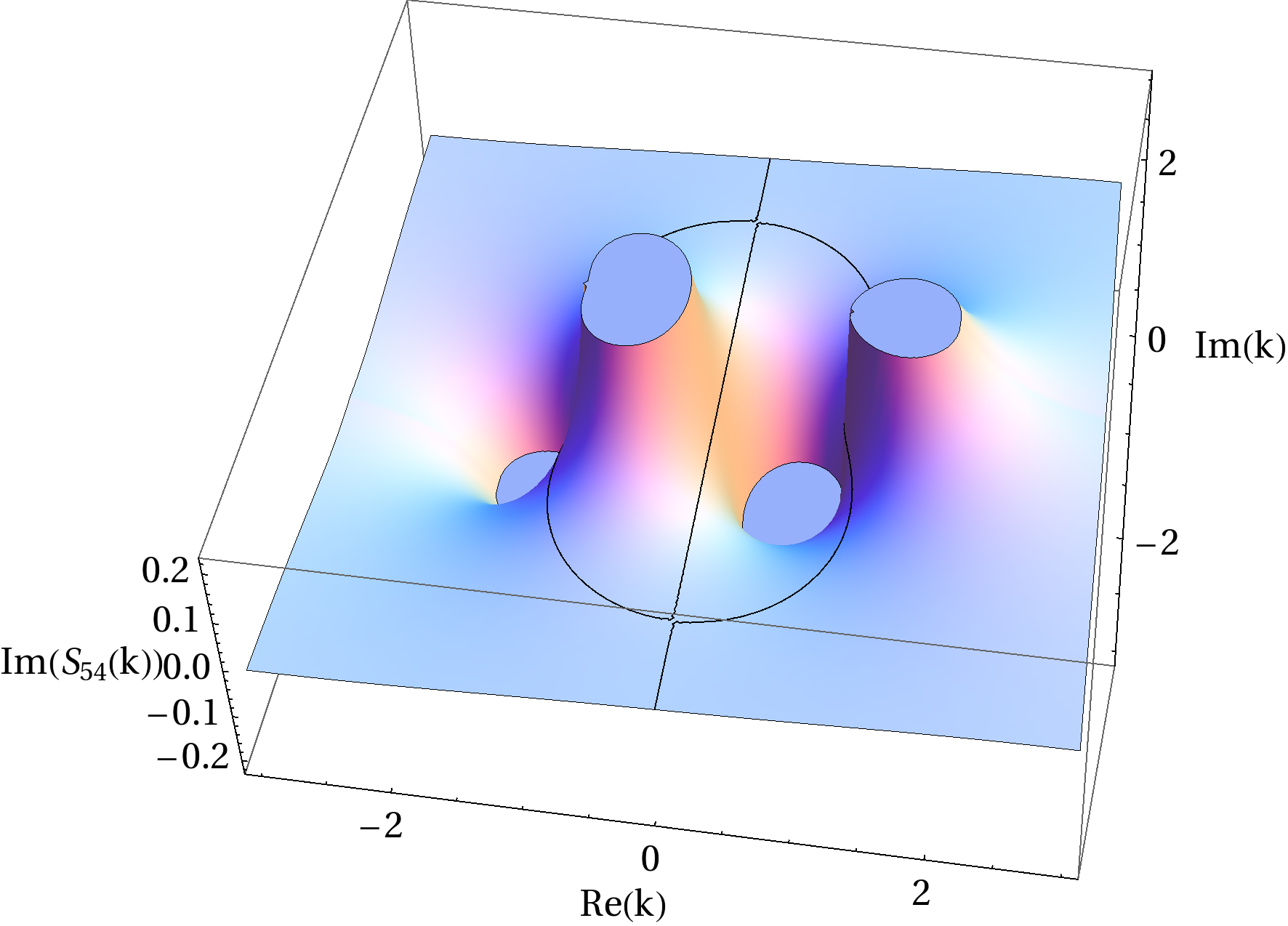}
\caption{Plot of $\IM S_{54}$ \eqref{eq:S54} with marked points at which the function vanishes.\label{fig:imS54}}\end{figure}

To determine the locus of `physical' $k$ on the complex plane, we need to turn to the reality conditions following from the \slc\ representation theory, that is
\begin{align} 
S_{54}=2\nu\in\mathbb R 
\end{align}
This condition is met for purely imaginary $k$, which is obvious when one notes that then 
$K(k^2),E(k^2)\in\mathbb R$. This is, however, not the most general condition, as evidenced in a plot of $\IM S_{54}$ (see \figref{fig:imS54}). Indeed there exist curved paths of physical $k$ that branch from the purely imaginary axis into the complex plane and terminate at the singularities at $k=\pm1$. 

Let us now describe in more detail the behaviour of $\nu$. On the imaginary axis in the
upper half plane $\nu$ initialy rises from 0, reaches a maximum value $\nu_*$ and then
falls off to a constant ($\sqrt{\lm}/\pi$) when $k \to +i\infty$. The intercept is
real and negative, has a minimum at $\nu=\nu_*$ and approches 0 as $k \to +i\infty$.
Of course, from the point of view of \slc\ representations we do not expect to have any
bound on $\nu$. And indeed, as shown in \figref{fig:imS54}, we find that real $\nu>\nu_*$
are realized on a curved path of complex $k$. Surprisingly, the intercept acquires
an imaginary part there. Nevertheless, in what follows we will work
only with purely imaginary $k$ as we are mainly interested in smaller values
of the intercept. However we find the whole structure and behaviour of the classical
strong coupling BFKL intercept unexpectedly subtle.

\section{Algebraic curves and particle interpretation}

The current paradigm of integrability for the spectral problem can be understood
as a progression of three basic stages. Firstly, one can analyze a particular operator/string state in terms of the Asymptotic Bethe Ansatz. At this stage one can identify the 
momentum carrying Bethe roots with specific particle content of the particular
excited state of the string. In the second stage, one can add the contribution
of leading wrapping corrections through L{\"u}scher formulas. In these formulas,
the particle content is still explicit. Finally, in the exact TBA/FiNLIE formulations
which resum all multiple wrapping terms, the notion of particle content still
plays a fundamental role as specifying driving terms in the integral equations
(or equivalently some specific analytical properties of the relevant functions entering
the equations).

For this reason, the question of particle content of a particular string solution
becomes very important. However this is especially mysterious if one analyses
BFKL physics. There, there are two continuous charges which appear naturally -- $j$ and
$\nu$. It is very difficult to imagine how this can be realized in terms of 
a multi-particle state with a definite (integer) number of excitations characteristic
of an integrable worldsheet theory.

In fact, the above puzzle was the main motivation for undertaking the present investigation.
Of course, the notion of deducing the multiparticle state corresponding to a given
classical solution is far from trivial. For the question even to make sense,
we have to deal with an integrable QFT. Fortunately this is the case here.
However, since the theory is nonlinear, extracting the particle picture is still
not trivial. The way this can be done is to use the algebraic curve construction
of classical string solutions which can be viewed as some kind of `nonlinear Fourier
transform' of the classical solution.

Within the algebraic curve construction (also known as finite-gap construction),
the branch cuts of the algebraic curve correspond to the location of Bethe roots of the
Asymptotic Bethe Ansatz, which in turn correspond to excitations of the string.
Moreover, we can identify the momenta and energies of elementary string excitations forming the classical string solution. In the next section we will show how the `BFKL folded string'
solution found earlier can be embedded into the algebraic curve construction.
We will then identify the branch cuts and energies and momenta of the constituents.

Before turning to the concrete examples, let us make here some general comments. First, we will recall --- and adapt to our setting --- the introduction of the particle density. 
This is of course very well known (see \cite{KMMZ,KZ,AFS}), but due to some subtle
differences (like lack of any charge on the $S^5$) it is convenient for us to review
the formulas here.

Let us start with the Bethe ansatz equations, given here for the \slii sector \cite{BS05}:
\begin{align}
\left(\frac{x_k^+}{x_k^-}\right)^J=\prod^M_{\substack{j=1\\j\neq k}}\left(\frac{x_k^+-x_j^-}{x_k^--x_j^+}\right)^{-1}\frac{1-\frac{g^2}{x_k^+x_j^-}}{1-\frac{g^2}{x_k^-x_j^+}}\sigma^2(x_k,x_j)
\end{align}
with the usual shorthands of $x_j=x(u_j),x_j^\pm=x(u_j\pm\frac i2)$, where $\sigma$ is the dressing phase and the spectral variable $x$ is related to the Bethe roots by the Zhukovsky transformation
\begin{align}
u(x)=x+\frac{g^2}{x}
\end{align}
In the typical setting, the roots are rescaled with $J$, but in our case this charge equals 0 from the outset (so the left-hand side is trivial). Instead, as we work in the strong coupling limit, the variables are rescaled by $g$. The leading order of the logarithm of the Bethe equation reads
\begin{align}
2i\pi n=\frac ig\sum^K_{\substack{j=1\\j\neq k}}\frac{-2x_kx_j(x_kx_j-1)}{(x_k^2-1)(x_j^2-1)(x_k-x_j)}
\end{align}
or, in the continuum limit, integrating along all cuts,
\begin{align}
\pi n=\frac1g\dashint\frac{-xx'(xx'-1)\rho(u')du'}{(x^2-1)(x'^2-1)(x-x')}
\label{eq:BethePV}
\end{align}
where $du'=(1-\frac1{(x')^2})dx'$ and we have introduced a density of roots in the $u$ variable.
For cuts symmetric under $x \to -x$, the above equation simplifies to give
\eq
\pi n = -\f{1}{g} \dashint \f{2 \rho(u') x dx'}{x^2-{x'}^2}
\label{eq:BethePVsym}
\eqx
In the above formulas, the density $\rho(u)$ is normalized in the following way:
\eq
\label{e.rhonorm}
\int \rho(u) du=K
\eqx
where the number of roots $K$ coincides with the spin. Care is necessary when the
roots are complex. Then obviously $\rho(u)du$ is real (and positive) along the cut.

Let us now compare the preceding formulas to the ones following from the algebraic curve
construction. The basic algebraic curve approach does not determine the exact location of the cuts, except of course for their endpoints. The unimodularity condition
\begin{align}
p(x-\epsilon)+p(x+\epsilon)=2n\pi
\end{align}
does not fix it either: we are free to flip the sign of the square root of $dp$ over some area, i.e.\ partially bordered by the cut. In order to determine the position of the cut, one identifies
the algebraic curve counterpart of (\ref{e.rhonorm}) and imposes the condition that
the resulting $\rho(u)du$ is \emph{real} along the cut.

Using the asymptotics of the pseudomomentum \cite{KZ}
\begin{align}
p(x\approx0) = \frac{2\pi(S-E)}{\sqrt\lambda} x + \dotsb \qqqq p(x\approx\infty) = \frac{2\pi(S+E)}{\sqrt\lambda x} + \dotsb
\label{eq:pAsympCharges}
\end{align}
and integrating $p(x)du$ anticlockwise around the cuts
\eq
\oint_C p(x) du =\oint_C p(x) \left(1-\f{1}{x^2}\right) dx 
\eqx 
and deforming the contour $C$ to encircle $x=0$ and $x=\infty$ one obtains
\begin{align}
\int\disc p\,du&=\left(\oint_0+\oint_\infty\right)p(x)\left(1-\frac1{x^2}\right)dx\notag\\&=2\pi i(-2\pi(S-E)-2\pi(S+E))/\sqrt\lambda=-8\pi^2iS/\sqrt\lambda
\end{align}
Comparing the above equation with (\ref{e.rhonorm}), one can identify the discontinuity of $p$
with the particle density in the $u$ variable:
\begin{align}
\rho=\frac{\sqrt\lambda}{-8\pi^2i}\disc p
\label{eq:densitydef}
\end{align}
The location of the cuts will then follow from the reality of $\rho(u)du$ defined in the above
way.

For completeness, let us recall how to determine the energies and momenta of the constituent
particles. Under assumption that the momenta scale as $p=P/g$, the leading $g\to\infty$ term gives the following relation:
\begin{align}
u=\frac{\sqrt{1+4P^2}}P
\end{align}
leading to the expression for the momentum
\eq
P(x)=\f{x}{x^2-1}
\eqx
Inserting the result in the dispersion relation
\begin{align}
\epsilon(p)=\sqrt{1+16g^2\sin^2\tfrac p2}
\end{align}
gives the magnon energy in the above limit
\begin{align}
\epsilon(x)=\frac{x^2-1}{x^2+1}
\end{align}
Indeed, using the asymptotic properties (\ref{eq:pAsympCharges}) one easily verifies that
\begin{align}
\int \rho(x) \epsilon(x)\,du=\int\rho(x)\frac{x^2+1}{x^2}dx=E
\label{eq:energynorm}
\end{align}
again integrating along all cuts.

\section{Algebraic curve analysis}

The aim of this section is to identify the particle content of the BFKL folded string
from the algebraic curve construction. In particular we would like to find the location
of the branch cuts for physical values of $k$ determined in section~\ref{s.class} (at
least for $\nu<\nu_*$). We will start by discussing the case of the 
GKP folded string, where the answer is well known, before proceeding to study the BFKL
folded string. 

\subsection{The GKP folded string}

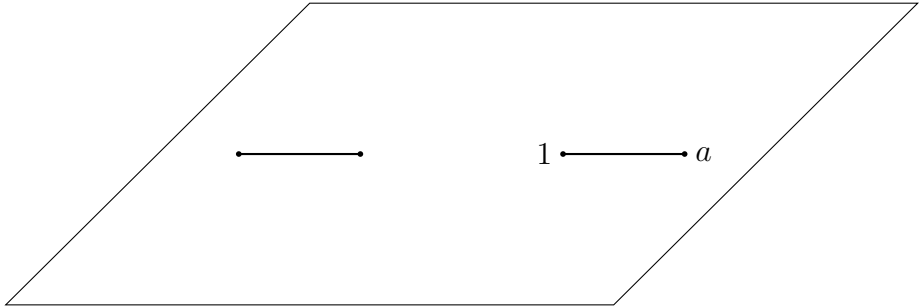
\begin{figure}\begin{center}
\begin{tikzpicture}[scale=4/3]
\begin{scope}[xslant=1,yscale=0.5]
    \draw (-3,-3) rectangle (3,3);
    \begin{scope}[every node/.style={fill,circle,inner sep=0.75}]
        \draw[thick] (0:1) node{} -- (0:2.2) node{};
        \draw[thick] (180:1) node{} -- (180:2.2) node{};
    \end{scope}
    \draw (0:1) node[left]{$1$};
    \draw (0:2.2) node[right]{$a$};
\end{scope}
\end{tikzpicture}
\end{center}\caption{Position of cuts for the GKP regime, $a$ is real and larger than 1.}\label{f:gkpcuts}\end{figure}

As argued in \cite{JLg12}, the GKP folded string is described by the following 
quasi-momentum\footnote{Note that here $J=0$. Although the conserved charges are smooth in $J$, the pseudomomentum $dp$ cannot be obtained
directly as a limit $J \to 0$ of the known $J>0$ expression.}:
\begin{align} 
dp = \frac{Ax^2+B}{(x^2-1)\sqrt{(x^2-1)(x^2-a^2)}}dx 
\end{align}
The cuts between the sheets extends between 1 and (real) $a$, as well as between $-1,-a$ (see \figref{f:gkpcuts}), and the energy and spin can be extracted from the asymptotic behaviour described in \cite{KZ},
which in this case amounts to 
\begin{align} 
p(x\approx0) = -\frac Ba x + \dotsb \qqqq p(x\approx\infty) = \frac Ax + \dotsb 
\label{eq:pAsympSigns}
\end{align}
%\begin{align} p(x\approx0) &= -\frac Ba x + \dotsb && \Rightarrow & \frac Ba &= 2\pi(E-S) \label{eq:pAsympZero}\\
%         p(x\approx\infty) &= \frac Ax + \dotsb && \Rightarrow & A &= 2\pi(E+S) \label{eq:pAsympInfty}\end{align}

The parameters $A,B$ can be fixed by computing the A-cycle, B-cycle, and $\Gamma$-contour (see \cite{KZ}) integrals of $dp$, which, for any algebraic curve, are respectively defined along contours encircling any given cut, linking two cuts, and linking a cut with infinity. In our case, which consists of just two cuts, the A-cycle can be deformed to a contour along the imaginary axis closed off by an infinite semi-circle at either side. The integral corresponding to the latter part of the contour vanishes, therefore the A-period is equal 
to the integral of $dp$ along the imaginary axis that evaluates to
\begin{align} 
\oint_A dp = \frac{2i}{a^2-1}\left(-(A+B)E(1-a^2)+(a^2A+B)K(1-a^2)\right) 
\label{eq:Acycle}
\end{align}
This expression is required to vanish, yielding an expression for one of the coefficients.

%\begin{align} B=A\frac{a^2K(1-a^2)-E(1-a^2)}{E(1-a^2)-K(1-a^2)} \label{eq:ABrelation}\end{align}

The B-cycle can be deformed to pass through infinity, therefore splitting to a sum of two 
$\Gamma$-contours. They in turn will be identical due to evenness of $dp$. We need to evaluate just one of them, and we start with the result of symbolic integration of $dp$:
\begin{align} 
p(x) &= -\frac{(A+B)x}{a^2-1}\cdot\frac{\sqrt{x^2-a^2}}{\sqrt{x^2-1}} \notag\\
                   &+ \tfrac1{a(a^2-1)}\left(a^2(A+B)E(\arcsin x\vert\tfrac1{a^2})+(1-a^2)BF(\arcsin x\vert\tfrac1{a^2})\right)
\label{eq:explicitP}
\end{align}
The $\Gamma$-contour is now equal to the value of this function at any point of the cut, and it is quite easy to manipulate at $x=a$, where the first term vanishes. After applying several identities for elliptic integrals, most notably the Legendre relation (see \cite{HMF10}), as well as the condition obtained from vanishing of \eqref{eq:Acycle}, we arrive at the result 
\begin{align} 
\int_\Gamma dp = p(a) = \frac\pi2\cdot\frac A {E(1-a^2)-K(1-a^2)} = \pi n 
\end{align}
This equation, and the vanishing of the A-period, allows us to fix both coefficients $A$ and $B$ in terms of the position
of the branch point $a$. We obtain 
\begin{align} 
E &= %\frac n{2\pi} \frac{a-1}a (E(1-a^2)+aK(1-a^2)) =
                   \sqrt\lambda\frac n{2\pi} \frac{a-1}a \left(aE(1-\tfrac1{a^2})+K(1-\tfrac1{a^2})\right) \\
              S &= %\frac n{2\pi} \frac{a+1}a (E(1-a^2)-aK(1-a^2)) =
                   \sqrt\lambda\frac n{2\pi} \frac{a+1}a \left(aE(1-\tfrac1{a^2})-K(1-\tfrac1{a^2})\right) 
\end{align}
reproducing a special case of the expressions deduced in \cite{GS$^+$11}. These formulas
allow us to link the algebraic curve parameter $a$ (location of the branch point)
with the parameter $k$ appearing in the classical solution of section~\ref{s.class}, in terms
of which we have expressed the physicality conditions.
Indeed these formulas can be matched (for $n=1$) with \eqref{eq:gkpE}-\eqref{eq:gkpS} 
using a (descending) Landen transformation
\begin{align}
a=\frac{1+k}{1-k}
\end{align}
Consequently, since in this case $k\in (0,1)$, $a$ is a real number greater than~1. It is
straightforward to check that the reality condition for $\rho(u) du$ are satisfied
for the cuts being the interval $(1,a)$ and its mirror image.

\subsection{The BFKL folded string}

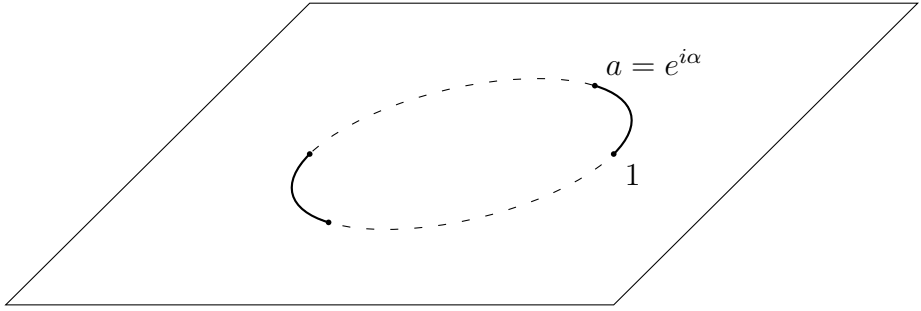
\begin{figure}\begin{center}
\begin{tikzpicture}[scale=2]
\begin{scope}[xslant=1,yscale=0.5]
    \draw (-2,-2) rectangle (2,2);
    \draw[loosely dashed] (0,0) circle (1);
    \begin{scope}[every node/.style={fill,circle,inner sep=0.75}]
        \draw[thick] (0:1) node{} arc (0:65:1) node{};
        \draw[thick] (180:1) node{} arc (180:245:1) node{};
    \end{scope}
    \draw (0:1) node[below right]{$1$};
    \draw (65:1) node[above right]{$a=e^{i\alpha}$};
\end{scope}
\end{tikzpicture}
\end{center}\caption{Position of cuts for the BFKL regime, $a$ lies on the unit circle.}\label{f:bfklcuts}\end{figure}

Since the BFKL folded string amounts just to a reshuffling and complexification of
the embedding coordinates of the GKP folded string, the preceeding discussion
will follow through, together with the relation for the branch point
\begin{align}
a=\frac{1+k}{1-k}
\end{align}
Now, however, the physical values of $k$ are (for $0\leq\nu\leq \nu_*$) purely imaginary and thus
$a$ will have unit modulus\footnote{A convenient parametrization is $k=i\tan \alpha/2$, which leads to $a=e^{i\alpha}$.}, allowing a guess that in this case the cuts will lie on the unit circle, as shown in \figref{f:bfklcuts}.

To verify this, we need to impose reality conditions on the density $\rho(u) du$. Indeed
one can check that on the unit circle $\disc p$ is purely imaginary, while $du$ is
obviously real (parametrizing the unit circle as $x=e^{i\phi}$) thus satisfying 
the reality condition. Note that this location of the branch cut leads to purely
imaginary energies and momenta of the constituent magnons.

\begin{figure}[h]\begin{center}
\includegraphics[width=0.7\textwidth]{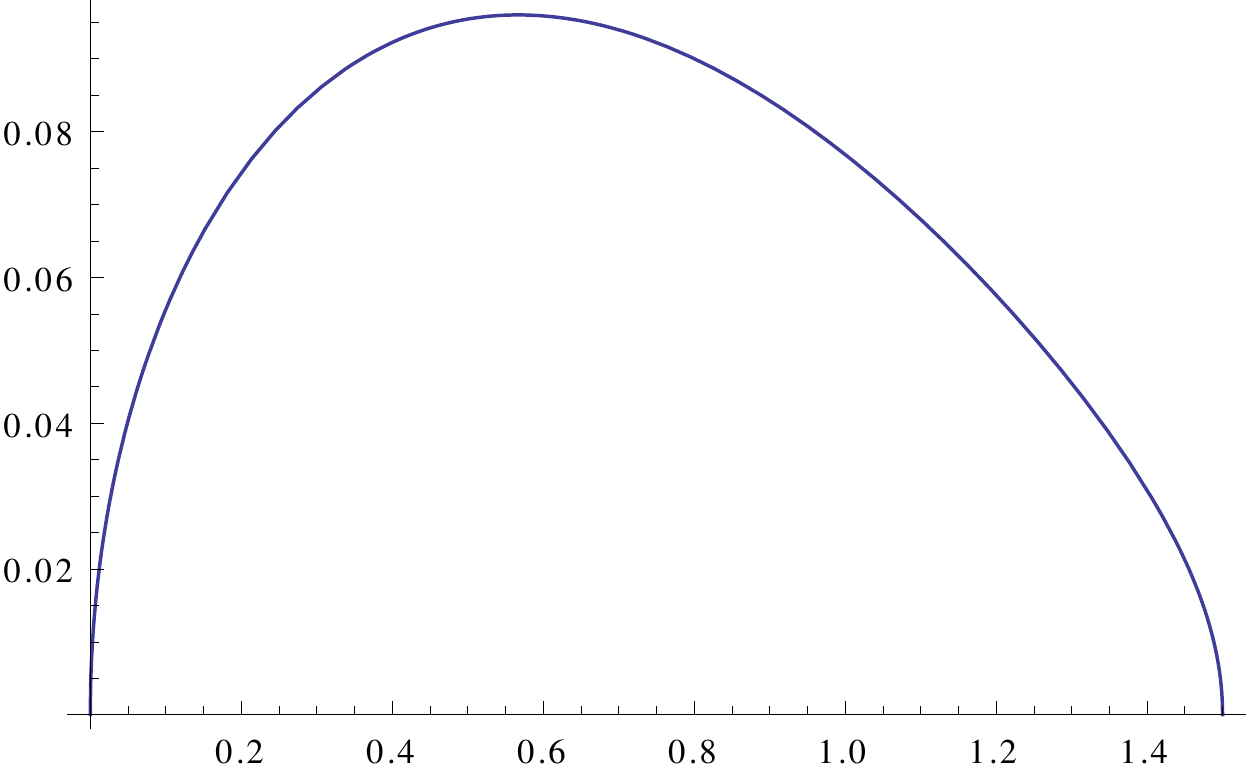}
\end{center}\caption{The integrand of \eqref{eq:BFKLdensitynorm}, i.e.\ $\rho(e^{i\phi})\frac{du}{d\phi}/\sqrt\lambda$, for $\alpha=\frac32$ and $\phi\in[0,\alpha]$.}
\label{f:integrandJ}\end{figure}

Let us finally express $\nu$ and the intercept $j$ in terms of integrals of some real density. 
By comparing pairwise \eqref{eq:gkpE}-\eqref{eq:gkpS} and \eqref{eq:S54}-\eqref{eq:S03}, we can reformulate the $p$ asymptotics as follows:
\begin{align}p(x\approx0)=\frac{2\pi(j-2i\nu)}{\sqrt\lambda}x+\dotsb\qquad p(x\approx\infty)=\frac{2\pi(j+2i\nu)}{\sqrt\lambda x}+\dotsb\end{align}
The density integrals evaluate thus to
\begin{align}\frac{\sqrt\lambda}{-8\pi^2i}\int\disc p\,du=-\frac{\sqrt\lambda}{2\pi^2}\int_0^\alpha\IM p(e^{i\phi})(1-e^{-2i\phi})ie^{i\phi}d\phi&=j\in\mathbb R\label{eq:BFKLdensitynorm}\\
\frac{\sqrt\lambda}{-8\pi^2i}\int\disc p\,\epsilon\,du=-\frac{\sqrt\lambda}{2\pi^2}\int_0^\alpha\IM p(e^{i\phi})(1+e^{-2i\phi})ie^{i\phi}d\phi&=2i\nu\in i\mathbb R\end{align}
The integrand of the former is plotted in \figref{f:integrandJ}.

One can also numerically verify the continuum Bethe equations \eqref{eq:BethePVsym}. 
Thus on the classical level there does not seem to be a problem with a particle
description of the BFKL folded string solution. There is just one subtlety.
The normalization of the density to the spin as in \eqref{e.rhonorm} suggests that the particle density is in this case negative definite.
This might indicate a need for some kind of continuum Baxter equation interpretation, 
however the difference seems so minor that it could be undone by redefining
the quantum numbers carried by elementary magnons (recall that we are not in the
standard \slii sector now but in a reshuffled/complexified version).
Classical considerations are not enough to unambigously settle this issue.

\section{Conclusions}

In this paper we have used the identification between quantum numbers of 
the BFKL pomeron and specific charges of the complexified conformal group
to bypass the neccessity of performing analytical continuation of
anomalous dimensions of specific operators in order to study high energy
scattering physics. 

In the strong coupling regime, we have identified classical string solutions
with the relevant conserved charges. We have restricted ourselves, for simplicity,
to the case of zero conformal spin where the results could be cross checked with
the results obtained using analytical continuation of anomalous dimensions
of twist-2 operators. However the same methods could be applied to the case
of nonzero conformal spin, where it is not known \emph{a~priori} 
what is the precise class of operators which would be linked by analytical 
continuation. In addition, other non-leading string solutions might be
the counterparts of compound multi-reggeized gluon states. For these
other states the link with anomalous dimensions of specific operators
is unknown as well. 

The relevant classical solution for vanishing conformal spin -- the BFKL 
folded string -- is a very close cousin of the well known GKP folded string.
It immediately reproduces the classical part of the strong coupling pomeron
intercept.
Our main focus, however, was on investigating the integrable properties of this
solution and its particle-like description. We have analyzed the algebraic
curve construction of this solution and identified the locations of the cuts
which, for a wide range of parameters, lie on two arcs on the unit circle.
The solution is consistent with a real density of particles (at the classical
level, its sign could be perhaps absorbed in a redefinition of the charges of the
elementary magnons), so at the classical level the integrable description
of this solution does not exibit any marked pathologies.
This is in contrast to the weak coupling case, where the Baxter equation 
description is very nonstandard~\cite{RJBAXTER}.

There are numerous avenues for further research. Firstly, one could
investigate more general solutions with nonzero conformal spin and find
the classical intercept in these cases. Secondly, it would be very interesting
to perform a short string limit along the lines of \cite{GS$^+$11,RT}.
Finally, the direct algebraic curve description of a classical solution
relevant for BFKL might be useful in formulating a Y-system approach.

\paragraph{Acknowledgments}
This work is supported by the International PhD Proj\-ects Programme of the Foundation for Polish Science within the European Regional Development Fund of the European Union, agreement no.\ MPD\hspace{0pt}/\hspace{0pt}2009\hspace{0pt}/\hspace{0pt}6. RJ was supported by NCN grant 2012/06/A/ST2/00396. PLG thanks the Laboratory of Theoretical Physics of École Normale Superieure in Paris for hospitality during the period when a part of this work has been performed. RJ thanks IPhT Saclay for hospitality when this work
was finalized.

%%%%%%%%%%%%%%%%%%%%% IAS?

\appendix

\section{Notation}

$F,K$, and $E$ are the standard elliptic integrals. For complete integrals, we use the `Mathematica' notation, in which the argument is squared with respect to the most traditional one, i.e.\ $E(k^2)=E(\frac\pi2\vert k^2)$. $\cn$ is one of the Jacobi elliptic functions.

\end{document}